\documentclass[10pt]{iopart}
\usepackage{iopams}  
\usepackage[intlimits]{esint}
\usepackage{bm,latexsym,mathrsfs,enumerate,bbm}
\usepackage{temp}
\usepackage{upgreek}
\usepackage[table,x11names]{xcolor}
\usepackage[breaklinks=true,unicode=true,urlcolor = RoyalBlue4,colorlinks= true,citecolor = RoyalBlue4,linkcolor = RoyalBlue4]{hyperref}
\usepackage{graphicx}
\usepackage[cal=boondoxo,scr=rsfs]{mathalfa}
\renewcommand{\vec}[1]{\bm{#1}}
\newcommand{\py}{Ni$_{80}$Fe$_{20}$}
\usepackage{siunitx}
\newcommand{\ch}[1]{#1}
\newcommand{\BesselJ}{\mathrm{J}}
\newcommand{\BesselY}{\mathrm{Y}}
\newcommand{\ellipticE}{\mathrm{E}}
\newcommand{\ellipticK}{\mathrm{K}}
\usepackage[numbers,sort&compress]{natbib}
\usepackage{easyReview}

\begin{document}

\title[Circular stripe domains\ldots]{Circular stripe domains and cone state vortices in disk-shaped exchange coupled magnetic heterostructures}

\author{Oleksandr Zaiets$^{1,2,3}$, Volodymyr P. Kravchuk$^{4,5}$, Oleksandr V. Pylypovskyi$^{3,1}$, Denys Makarov$^3$, Denis D. Sheka$^6$}
%\ead{o.zaiets@kau.edu.ua}
\address{$^1$Kyiv Academic University, 03142 Kyiv, Ukraine}
\address{$^2$Leibniz Institute for Solid State and Materials Research, 01069 Dresden, Germany}
\address{$^3$Helmholtz-Zentrum Dresden-Rossendorf e.V., Institute of Ion Beam Physics and Materials Research, 01328 Dresden, Germany}

%\author{Volodymyr P. Kravchuk}
%\ead{volodymyr.kravchuk@kit.edu}
\address{$^4$Institut f\"{u}r Theoretische Festk\"{o}rperphysik, Karlsruher Institut f\"{u}r Technologie, 76131 Karlsruhe, Germany}
\address{$^5$Bogolyubov Institute for Theoretical Physics of National Academy of Sciences of Ukraine, 03143 Kyiv, Ukraine}

%\author{Oleksandr V. Pylypovskyi}
%\ead{o.pylypovskyi@hzdr.de}
%\address{Helmholtz-Zentrum Dresden-Rossendorf e.V., Institute of Ion Beam Physics and Materials Research, 01328 Dresden, Germany}
%\address{Kyiv Academic University, 03142 Kyiv, Ukraine}
%
%\author{Denys Makarov}
%\ead{d.makarov@hzdr.de}
%\address{Helmholtz-Zentrum Dresden-Rossendorf e.V., Institute of Ion Beam Physics and Materials Research, 01328 Dresden, Germany}
%
%\author{Denis D. Sheka}
\address{$^6$Taras Shevchenko National University of Kyiv, 01601 Kyiv, Ukraine}
\ead{sheka@knu.ua}

\vspace{10pt}
\begin{indented}
	\item[]\today
\end{indented}

\begin{abstract}
Vertically stacked exchange coupled magnetic heterostructures of cylindrical geometry can host complex noncolinear magnetization patterns. By tuning the interlayer exchange coupling between a layer accommodating magnetic vortex state and an out-of-plane magnetized layer, one can efficiently realize new topological chiral textures such as cone state vortices and circular stripe domains. We study how the number of circular stripes can be controlled by both the interlayer exchange coupling and the sample geometrical parameters. By varying geometrical parameters, a continuous phase transition between the homogeneous state, cone state vortex, circular stripe domains, and the imprinted vortex takes place, which is analysed by full scale micromagnetic simulations. The analytical description provides an intuitive pictures of the magnetization textures in each of these phases. The possibility to realize switching between different states allows for engineering magnetic textures with possible applications in spintronic devices.	
\end{abstract}

\vspace{2pc}

\submitto{\JPD}
\ioptwocol
%
%%%%%%%%%%%%%%%%%%%%%%%%%%%%%%%%%%%%%%%%%%%%%%%%%%
%%%			New section
%%%%%%%%%%%%%%%%%%%%%%%%%%%%%%%%%%%%%%%%%%%%%%%%%%

\section{Introduction}
\label{sec:intro}

Out-of-plane magnetized thin films can host different topologically protected magnetic textures, which are actively discussed for applications in information storage, magnetic random access memory, sensors, and neuromorphic computing \cite{Parkin15, Zang18, Grollier20}. These magnetic textures include topological magnetic solitons such as domain walls, skyrmions, bimerons, skyrmion-bubbles, skyrmionium \cite{Goebel21}. There are different ways to realise these objects. In particular, while chiral skyrmions can result from the Dzyaloshinskii--Moriya interaction \cite{Fert17,Wiesendanger16}, skyrmion-bubbles are stabilized by a nonlocal magnetostatic interaction \cite{Jiang15}. Alternatively, non-collinear magnetic textures can be imprinted relying on the indirect coupling between nanodisks hosting vortex state and out-of-plane magnetization \cite{Streubel15, Streubel15d}. By tuning the interlayer exchange coupling in the layer stack, e.g. by modifying the thickness of nonmagnetic spacer, topological textures in the out-of-plane magnetized layer can be efficiently manipulated. Still,  the description of non-collinear magnetic textures in coupled heterostructures is missing. 

Here, we develop a theoretical approach (analytics and simulations) for the description of different magnetization states in a coupled heterostructure consisting of an in-plane ferromagnet in a vortex state / nonmagnetic spacer / out-of-plane ferromagnet with various material and geometrical parameters. A phase diagram of equilibrium magnetization states in a disk is assembled in a wide range of the interlayer exchange coupling parameters for different thicknesses of the layer with out-of-plane easy axis of magnetization. We analyse the transition between different states in the out-of-plane magnetized layer when the coupling strength increases: (i) homogeneous state is realized when the coupling is absent (e.g., the case of a very thick spacer), (ii) cone state vortex and circular stripe domain patterns are observed for intermediate coupling strength, and (iii) imprinted vortex state for the case of strong coupling. In particular, for the case of intermediate coupling strength, the modulation instability of the cone state vortex triggered by the nonlocal magnetostatic interaction results in the realization of a stable equilibrium magnetization texture in the form of circular stripes domains. The underlying physics is found to be similar to the formation of straight stripe domains in thin out-of-plane magnetized films exposed to an in-plane magnetic field \cite{Hubert09}. The density of stripe domains is controlled by the coupling strength and the thickness of the out-of-plane magnetized layer. The variety of magnetization textures with well-defined phases corresponds to the experimentally observed vortex and donut states reported by Streubel \emph{et al} \cite{Streubel15, Streubel15d}. 

%%%%%%%%%%%%%%%%%%%%%%%%%%%%%%%%%%%%%%%%%%%%%%%%%%
%%%			New section
%%%%%%%%%%%%%%%%%%%%%%%%%%%%%%%%%%%%%%%%%%%%%%%%%%

\section{Micromagnetic simulations of the phase diagram of equilibrium states}
\label{sec:diagram}

%==================================================================/
\begin{figure}
	\includegraphics[width=\columnwidth]{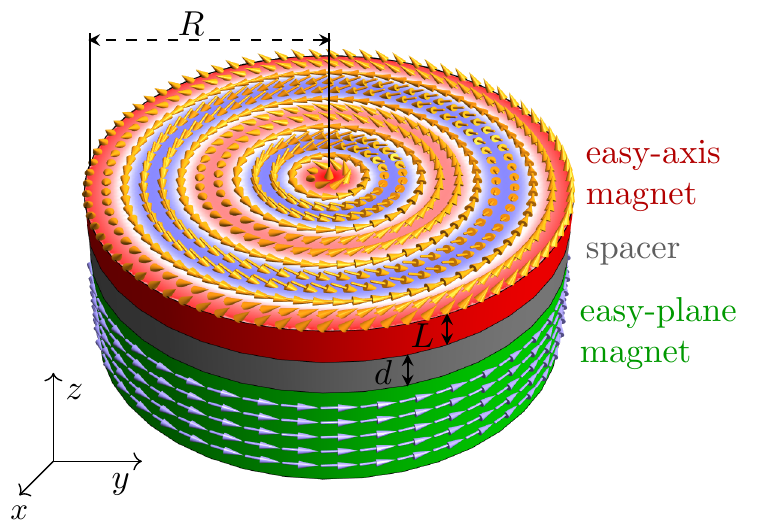}
	\caption{\textbf{A vertically stacked heterostructure}: schematics and notations. The heterostructure consists of 3 layers and is shaped as a disk of radius $R$. In the heterostructure, two magnetic layers are separated by a nonmagnetic spacer of thickness $d$. The layer with the out-of-plane easy axis of magnetization possesses thickness $L$. The magnetic state of this layer is affected by the magnetic texture in the bottom layer, which is fixed to be in the vortex state. Arrows show the local direction of magnetization in the top and bottom layers. Colorcode of the top surface indicates variation of the out-of-plane component of magnetization.}
	\label{fig:schematics}
\end{figure}
%==================================================================/

We consider a vertically stacked magnetic heterostructure with two magnetic layers separated by a nonmagnetic spacer, see Fig.~\ref{fig:schematics}. The heterostructure is shaped as a disk of radius $R$. The layer with in-plane magnetization is chosen to be appropriately thick to support the magnetic vortex state. The out-of-plane magnetized layer is of thickness $L$ with the perpendicular magnetized ground state. The thickness $d$ of the nonmagnetic spacer is adjusted to tune the interlayer exchange coupling between the two magnetic layers from strong (smaller $d$) to weak (larger $d$). Our model includes the following contributions to the energy of the out-of-plane magnetized layer: $E = E^{\text{x}} + E^{\textsc{a}} + E^{\text{c}} + E^{\textsc{ms}}$. Here, $E^{\text{x}} = -A\int \left(\vec{m}\cdot \nabla^2 \vec{m}\right) \rmd V$ describes the exchange contribution with $A$ being the exchange constant, $\vec{m}= \vec{M}/M_{\textsc{s}}$ is the normalized magnetization, $M_{\textsc{s}}$ is the saturation magnetization, and $V$ is the volume of the layer. The energy $E^{\textsc{a}} = -K \int \left(\vec{m}\cdot \hat{\vec{z}} \right)^2\rmd V$ with $K>0$ corresponds to the uniaxial anisotropy with the easy axis directed along $\hat{\vec{z}}$ direction. The energy of the interlayer exchange coupling is expressed as $E^{\textsc{c}} = -2\sigma \int \left(\vec{m}\cdot \vec{m}_{\text{fix}}\right) \rmd S$ with $\sigma$ being a biquadratic coupling coefficient, $\vec{m}_{\text{fix}}$ is the magnetization of the in-plane magnetized layer with fixed vortex state, and $S$ being the interface area \cite{Hubert09}. This coupling, which typically oscillates in sign as function of the thickness of the spacer layer, is closely related to the  Ruderman-Kittel-Kasuya-Yosida (RKKY) interaction between magnetic impurities, for a review see \cite{Stiles99}. The last energy term is the magnetostatic energy $E^{\textsc{ms}} = \frac{1}{2} M_{\textsc{s}}^2 \iint \rmd V \rmd V' \left(\vec{m}(\vec{r}) \cdot \vec{\nabla}\right) \left(\vec{m}(\vec{r}') \cdot \vec{\nabla}'\right) \left|\vec{r} - \vec{r}'\right|^{-1}$.

The ground state of the in-plane magnetized layer forms the magnetic vortex texture. Such a configuration is characterized by the absence of volume and edge surface magnetostatic charges. The only stray field comes from face surface charges, which are localized within the vortex core. Far from the vortex core, the magnetization can be described as
\begin{equation} \label{eq:m-fixed}
\vec{m}_{\text{fix}} = \mathcal{C}_{\text{fix}} \left(-\hat{\vec{x}}\sin\chi + \hat{\vec{y}}\cos\chi\right)	
\end{equation}
with $\mathcal{C}_{\text{fix}}=\pm1$ being the vortex circulation, which defines counter-clockwise or clockwise  direction of the magnetization, $\tan\chi=y/x$; $x$, $y$ are the coordinates in the disk plane. We assume that the magnetization state in the in-plane magnetized layer is not affected by the texture of the out-of-plane magnetized layer.

We study equilibrium magnetization states in the out-of-plane magnetized layer of the heterostructure. For this purpose, we perform a series of micromagnetic simulations in a wide range of the thickness of the out-of-plane magnetized layer $L\in[2,12]~\unit{nm}$ and coupling constant $\sigma\in[0,1.2]~\unit{mJ/m^2}$. For simulations, we use object oriented micromagnetic framework OOMMF \cite{OOMMFa,Donahue99}. The simulated heterostructure shaped as a disk of $R = 500~\unit{nm}$ included: (i) in-plane magnetized layer with parameters of Permalloy (\ch{Py}, \ch{\py}; exchange constant $A=13~\unit{pJ/m}$; saturation magnetization $M_\textsc{s} = 860~\unit{kA/m}$), (ii) out-of-plane magnetized layer with parameters typical for \ch{Co/Pt} multilayers  ($A=10~\unit{pJ/m}$, $M_\textsc{s} = 500~\unit{kA/m}$, $K = 200~\unit{kJ/m^3}$), which are separated by (iii) a nonmagnetic spacer of varying thickness $d$. Thermal effects are neglected in simulations. Mesh cells have size of $5\times 5\times 2~\unit{nm}^3$.
To decrease the simulation time, we pinned the magnetic texture of the in-plane magnetized layer in a vortex configuration with $\mathcal{C}_{\text{fix}}=+1$, using thin \ch{Py} layer of only $10~\unit{nm}$ thick.

%==================================================================\
\begin{figure*}
    \begin{center}
	\includegraphics[width=0.95\textwidth]{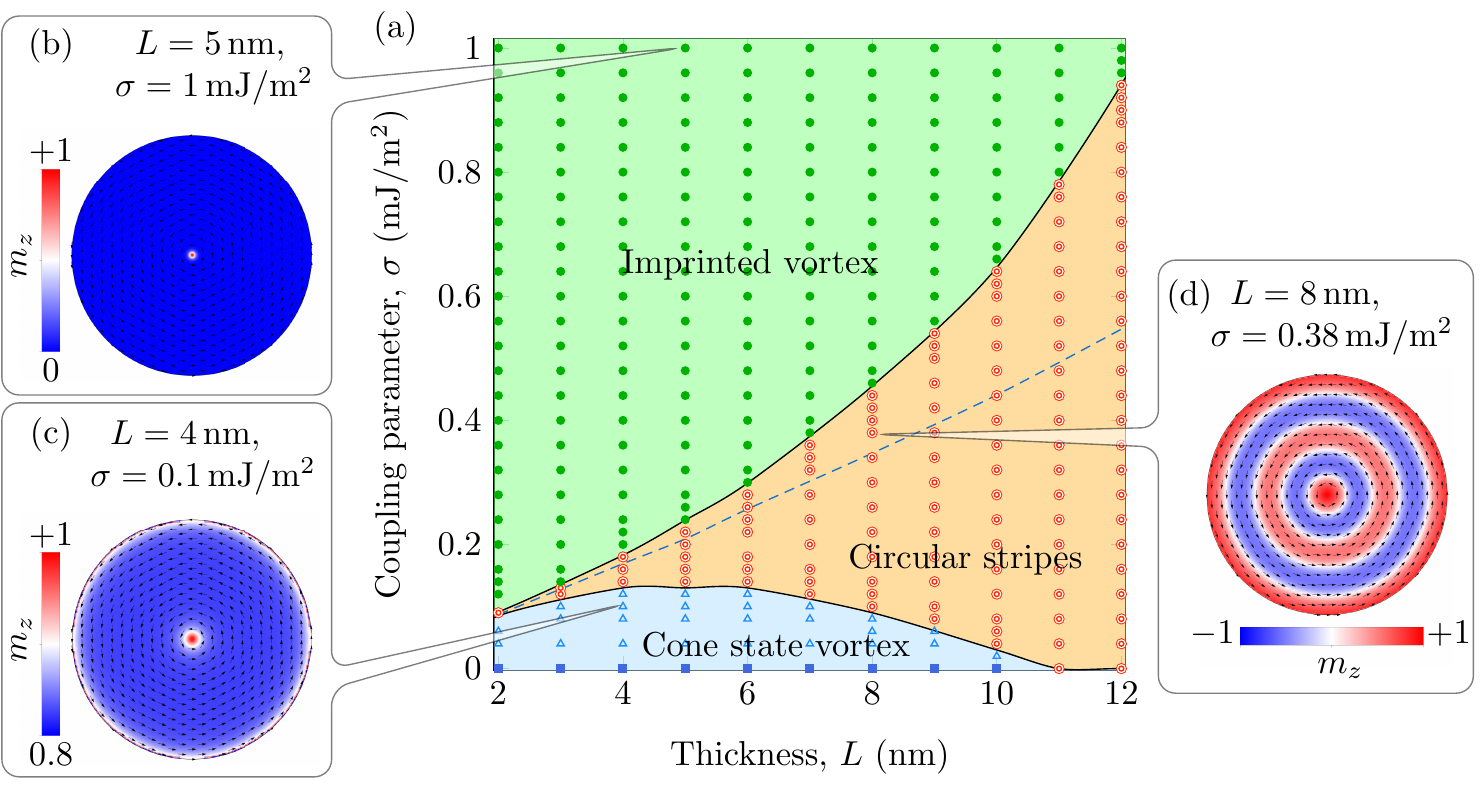}
    \end{center}
	\caption{\textbf{Equilibrium magnetization states.} (a) Phase diagram of equilibrium magnetization states for a disk with radius $R=500$~nm. Symbols correspond to OOMMF simulations with parameters summarized in~\ref{appendix:oommf_parameters}: homogeneous state is indicated with squares, light vortices in a cone state (triangles), circular stripes (rings), vortex state (filled circles). Panels (b)--(d) show the equilibrium magnetization texture for specified parameters. The dashed line describes the boundary of linear instability of the cone state vortex.}
	\label{fig:phasediagram}
\end{figure*}
%==================================================================/

Equilibrium states are determined using numerical energy minimization starting from two different configurations: saturated state along the anisotropy axis ($\hat{\vec{z}}$) and multidomain stripe state. By comparing energies of different states, we determine the energetically preferable states for different parameters: thickness of the out-of-plane layer $L$ and coupling parameter $\sigma$. Simulations data are shown in Fig.~\ref{fig:phasediagram}. When coupling is absent ($\sigma=0$), the free layer is homogeneously magnetized along the anisotropy axis with $m_z=\pm 1$. This phase is referred to as the \emph{homogeneous phase}. The ground state for the case of strong coupling is the \emph{imprinted vortex} state with the magnetization texture in the out-of-plane magnetized layer corresponding to the texture of the in-plane magnetized layer, $\vec{m}\approx  \vec{m}_{\text{fix}}$, see Fig.~\ref{fig:phasediagram}(a). If the coupling is not sufficiently strong to imprint the vortex, the radial symmetry of the texture remains but the perpendicular magnetization decreases with $|m_z|=\text{const}<1$. The magnetization in the ground state is directed along one of the direction of the cone with $m_z=\text{const}$ or the cone with $m_z=-\text{const}$. That is why we refer to this phase as a \emph{cone state vortex}, see Fig.~\ref{fig:phasediagram}(b). Such a state appears due to the competition between the easy-axis anisotropy and the interlayer exchange coupling of the out-of-plane magnetized layer to the layer hosting a vortex state, see Sec.~\ref{sec:cone-state}. Furthermore, another multidomain state appears for a moderate coupling strength between the imprinted vortex state and the cone state vortex. The magnetization texture resembles a set of homocentric rings with opposed perpendicular magnetization components, see Fig.~\ref{fig:phasediagram}(c). We will see below in Sec.~\ref{sec:circular-stripes} that this multidomain state is in many respects similar to straight stripe domains in perpendicular magnetized multilayers \cite{Suna86,Lemesh17}. Therefore, this phase is refereed to as a \emph{circular stripe domain}. 

%%%%%%%%%%%%%%%%%%%%%%%%%%%%%%%%%%%%%%%%%%%%%%%%%%
%%%			New section
%%%%%%%%%%%%%%%%%%%%%%%%%%%%%%%%%%%%%%%%%%%%%%%%%%
\section{Cone state vortex}
\label{sec:cone-state}

We put forth a model to gain insight to the mechanism of the formation of different states in a coupled heterostructure. We assume that the out-of-plane magnetized layer is sufficiently thin that the magnetization does not vary with thickness. With this, we limit the thickness $L$ not to exceed the effective magnetic length $\ell= \sqrt{A/K_{\text{ef}}}$ with $K_{\text{ef}} = K-2\pi M_{\textsc{s}}^2$. For typical parameters of \ch{Co/Pt} multilayers, $\ell\approx 15~\unit{nm}$. 
We also limit our consideration by the class of radially symmetric solutions:
\begin{equation} \label{eq:radial-symmetry}
\vec{m}(\vec{r}) = \hat{\vec{z}}\cos\theta(r) + \mathcal{C} \sin\theta(r) \left( -\hat{\vec{x}}\sin\chi + \hat{\vec{y}}\cos\chi \right),
\end{equation}
where the perpendicular magnetization component $m_z=\cos\theta(r)$ depends only on the polar radius $r=\sqrt{x^2+y^2}$, the parameter $\mathcal{C}=\pm1$ is the circulation of the magnetization texture. The symmetry of the magnetization texture \eqref{eq:radial-symmetry} is supported by the radial symmetry of the vortex magnetization $\vec{m}_{\text{fix}}$ of the in-plane magnetized layer. Besides, it corresponds to the absence of the volume magnetostatic charges in the out-of-plane magnetized layer. Both assumptions are well confirmed by micromagnetic simulations.

%%%%%%%%%%%%%%%%%%%%%%%%%%%%%%%%%%%%%%%%%%%%%%%%%%
%%%			New subsection
%%%%%%%%%%%%%%%%%%%%%%%%%%%%%%%%%%%%%%%%%%%%%%%%%%
\subsection{Model of a thin sample}
\label{sec:local}

We start our discussion with the model of a very thin out-of-plane magnetized layer, where the nonlocal magnetostatic energy of surface charges is replaced by the local anisotropy. This results in an effective easy axis anisotropy with the coefficient $K_{\text{ef}}>0$. While this model does not support stripe domains, it is important for the understanding of the structure of other states. The total energy of the model reads $E = E_0\int_0^1 \mathscr{E} \rho \rmd \rho$ with the energy density
\begin{equation} \label{eq:energy}
\mathscr{E} = \lambda^2 \left({\theta'}^2 + \frac{\sin^2\theta}{\rho^2}\right)	+\left(\sin\theta - h\right)^2.
\end{equation}
Here, $E_0 = 2\pi LR^2K_{\text{ef}}$ determines the energy of a uniformly magnetized sample, the parameter $\lambda = \sqrt{A/\left(R^2 K_{\text{ef}}\right)}$ determines the reduced length scale, and the prime denotes the derivative with respect to the reduced radial distance $\rho=r/R$. The last term in the energy functional \eqref{eq:energy} describes the interaction with an effective in-plane magnetic field. The amplitude of this effective field $h$ originates from the interlayer exchange coupling, namely $h=\sigma \mathcal{C}_{\text{fix}}\times \mathcal{C}/\left(L K_{\text{ef}}\right)$. To simplify the description, we suppose that the magnetization of the in-plane magnetized layer is characterized by the planar vortex structure \eqref{eq:m-fixed}. The interlayer exchange coupling results in the coupling between circulations of both magnetization textures, $\mathcal{C}_{\text{fix}}\times \mathcal{C} = \text{sgn }\sigma$: for the case under consideration with $\sigma>0$, the circulation of the equilibrium magnetization texture $\mathcal{C}$ coincides with the circulation of the vortex $\mathcal{C}_{\text{fix}}$ in the in-plane magnetized disk. Hence, we get the effective field $h=\sigma/\left(L K_{\text{ef}}\right)$.

%==================================================================/
\begin{figure*}
\includegraphics[width=\textwidth]{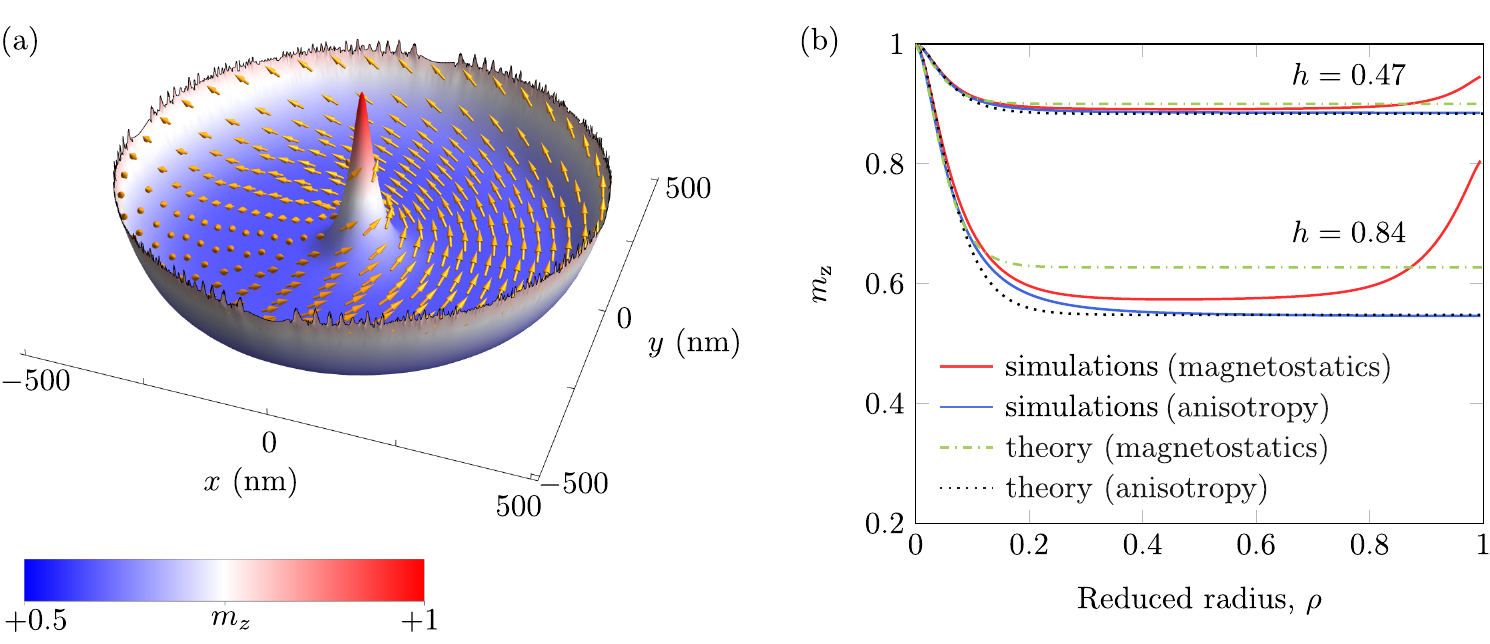}
\caption{ \label{fig:vortexincone}
\textbf{The magnetization texture of the cone state vortex}. (a) 3D representation of a light vortex texture from micromagnetic simulations of a heterostructure with a 5-nm-thick out-of-plane magnetized layer and a coupling strength $\sigma=0.18~\unit{mJ/m^2}$.  Other parameters are the same as in \ref{appendix:oommf_parameters}. (b) The profile of the perpendicular magnetization component along the radius of the disk. The data are shown for different coupling constants: $\sigma=0.1~\unit{mJ/m^2}$ ($h=0.47$) and $\sigma=0.18~\unit{mJ/m^2}$ ($h=0.84$). Red solid lines correspond to full scale micromagnetic simulations. Blue solid lines are micromagnetic simulations, where magnetostatic interaction is replaced with an effective anisotropy constant. Black dotted lines correspond to calculations within the theoretical two-parameter ansatz \eqref{eq:ansatz} with parameters \eqref{eq:mu-and-Delta-local} (local model). Green dash-dotted lines correspond to calculations within the theoretical ansatz \eqref{eq:ansatz} with parameters \eqref{eq:mu-and-Delta-nloc} (nonlocal model).
}
\end{figure*}
%==================================================================/

Without the effective magnetic field induced by the interlayer exchange coupling ($\sigma=0$, $h=0$), the equilibrium magnetization texture is homogeneous, $m_z=\pm1$. In the opposite case of a strong coupling, the layer with out-of-plane magnetization acquires the imprinted vortex state with $m_z=0$. Intermediate cases can be analysed far from the vortex core for $\rho \gg \lambda$:
\begin{equation} \label{eq:theta0}
m_z = \cases{\pm \sqrt{1-h^2}, & $h<1$\\ 0 & $h\ge1$.\\}
\end{equation}
The magnetization texture \eqref{eq:theta0} supplemented by the radially symmetric distribution \eqref{eq:radial-symmetry} corresponds to the magnetic vortex in the cone phase \cite{Kosevich83,Ivanov95b}. The magnetization direction in the centre of the vortex core being up or down, $m_z(0)=p\pm1$, determines the vortex polarity. In the following discussion, we consider vortices with $p=+1$, which is the same as the polarity of the vortex in the in-plane magnetized disk. Depending on the perpendicular magnetization component far from the vortex core, two types of vortices can be realized: a light vortex with $m_z=\sqrt{1-\mu^2}$ and a heavy vortex with $m_z=-\sqrt{1-\mu^2}$. We are interested in light vortices, which are energetically preferable, see discussion in 
\ref{sec:cone-state-appendix}. To describe the vortex structure affected by an effective magnetic field, we use a two-parameter ansatz 
\begin{equation} \label{eq:ansatz}
\sin\theta(\rho) = \mu \left[1-f\left(\frac{\rho}{\Delta}\right)\right],
\end{equation}
where the amplitude $\mu$ characterizes the absolute value of the equilibrium in-plane magnetization and the width $\Delta$ determines a typical width of the vortex core. To describe the vortex core profile, we use the exponentially localized function $f(\xi)=e^{-\xi}$, which is analogous to the Feldkeller ansatz \cite{Feldtkeller65}. The ground state of the model can be found by the minimization of energy with respect to variational parameters, which results in $\Delta\approx \lambda\sqrt2$ and $\mu\approx h$, see \ref{sec:cone-state-appendix} for details. The magnetization angle $\theta$ out of the vortex core is determined by $\sin\theta_0=\mu\approx h$. We note that the core size of the cone state vortex is almost independent of the amplitude of the effective field \cite{Note2}. This behaviour is distinct to the cone phase in easy-plane magnets, where the core of a light vortex spreads with a perpendicular field \cite{Kosevich83,Ivanov95b}.

%%%%%%%%%%%%%%%%%%%%%%%%%%%%%%%%%%%%%%%%%%%%%%%%%%
%%%			New subsection
%%%%%%%%%%%%%%%%%%%%%%%%%%%%%%%%%%%%%%%%%%%%%%%%%%

\subsection{Magnetization textures within the nonlocal model}
\label{sec:nonlocal}

In the framework of the local model valid for ultrathin samples, we identified three equilibrium magnetization states: homogeneous state, cone state vortex and imprinted vortex state. The existence of these states is confirmed by micromagnetic simulations, see Fig.~\ref{fig:phasediagram}. However, this model does not describe circular stripe domains. Moreover, the thickness of the out-of-plane magnetized layer $L$ does not enter the model \eqref{eq:energy}. To overcome these limitations, we extend our model to include the effects stemming from the nonlocal magnetostatic interaction and analyze their influence on the formation of magnetization textures in a coupled heterostructure. The radial symmetry \eqref{eq:radial-symmetry} of the magnetization texture allows to avoid volume magnetostatic charges. Hence, surface magnetostatic charges remain the only source of the demagnetization fields.

We start from the cone state vortex using the variational approach described above together with the ansatz \eqref{eq:ansatz}. Taking into account the energy of surface magnetostatic charges, the equilibrium values of the width of the vortex core $\Delta$ and the amplitude $\mu$ read
\begin{equation} \label{eq:mu-and-Delta}
\eqalign{\Delta\approx \frac{\lambda \sqrt{2}}{\sqrt{1 + \frac{8g(\varepsilon)}{Q-1} } },\quad \mu \approx \frac{h}{1 + \frac{g(\varepsilon)}{Q-1} },}	
\end{equation}
where $\varepsilon=L/(2R)$ is the sample aspect ratio, $Q=K/\left(2\pi M_{\textsc{s}}^2 \right)$ is the quality factor, and $g(\varepsilon)$ is defined in \eqref{eq:g-vs-epsilon}, see \eqref{eq:mu-and-Delta-nloc} for more details. The function $g(\varepsilon)$ vanishes for thin samples, which corresponds to the local model discussed above. In the opposite case of thick sample, $g(\varepsilon)$ tends to 1 resulting in a modification of the equilibrium parameters. 

A typical magnetization texture of a light vortex is shown in Fig.~\ref{fig:vortexincone} featuring a pronounced vortex core and the flux-free vortex configuration far from the core. The perpendicular magnetization component $\cos\theta$ is shown for two different values of an effective magnetic field $h$. We note that the ansatz \eqref{eq:ansatz} with the equilibrium parameters \eqref{eq:mu-and-Delta}, which take into account nonlocal magnetostatics, describes the simulations data well. However, it does not capture an increase of the magnetization at the edge of the disk observed in simulations. To understand the origin of this edge effect, we performed micromagnetic simulations, where the nonlocal magnetostatics was replaced by an effective anisotropy. In this case, far from the vortex core the magnetization tends to the the equilibrium value. Thus, the edge effect originates from the nonlocal influence of the surface magnetostatic charges.

We note that in our system cone state vortices are realized in magnets with perpendicular easy-axis anisotropy exposed to an effective in-plane magnetic field. It is instructive to mention an important difference with easy-plane magnets, where heavy and light vortices are separated by a large energy barrier (infinite barrier within the continuum model). The switching between these two states involves the appearance of a Bloch point \cite{Thiaville03,Kravchuk07a}. In contrast, light and heavy vortices in an easy-axis magnet possess comparable energies and can be transformed in one another by stray fields induced by surface magnetostatic charges. This effect leads to the formation of circular stripe domains.

%%%%%%%%%%%%%%%%%%%%%%%%%%%%%%%%%%%%%%%%%%%%%%%%%%
%%%			New subsection
%%%%%%%%%%%%%%%%%%%%%%%%%%%%%%%%%%%%%%%%%%%%%%%%%%

\section{Circular stripe domains}
\label{sec:circular-stripes}

Our further analysis is based on the theory of straight stripe domains, which are realized in films with sufficiently strong out-of-plane easy-axis anisotropy, see \ref{sec:straight-stripes}. By applying an in-plane magnetic field $H$, the magnetization can be tilted resulting in the appearance of a bistable state with $m_z\propto \pm \sqrt{4K_{\text{ef}}^2/M_{\textsc{s}}^2-H^2}$, see \eqref{eq:m0} for details. In a specific range of parameters (film thickness $L$, amplitude of the magnetic field $H$), this state can experience a modulation instability, which results in the formation of straight stripe domains \cite{Hubert09}. The density of straight stripe domains is proportional to $H^2 L$ for thin films and $H^{2/3}L^{-1/3}$ for thick films, see \ref{sec:straight-stripes} for details.

We exploit the analogy between the formation of straight stripe domains in a film exposed to an in-plane magnetic field $H$ and stabilization of circular stripe domains in coupled heterostructures driven by an effective in-plane field $h$ stemming from the interlayer exchange coupling. We note that a uniform magnetization texture in out-of-plane magnetized thin films becomes unstable in an in-plane field. Typically, the phase diagram consists of three states: (i) homogeneous titled state is realized in relatively weak fields; (ii) uniform in-plane state appears for strong fields, and (iii) straight stripe domains are realized when the strength of the in-plane field is moderate. Similarly, three states are realized in coupled heterostructures. The comparison of phase diagrams for disk-shaped coupled heterostructures and thin films is discussed in \ref{appendix:circular}. A clear similarity of the phase diagrams suggests the applicability of the theory of straight stripe domains for the description of circular stripe domains.

%%==================================================================/
\begin{figure*}
\begin{center}
\includegraphics[width=0.9\textwidth]{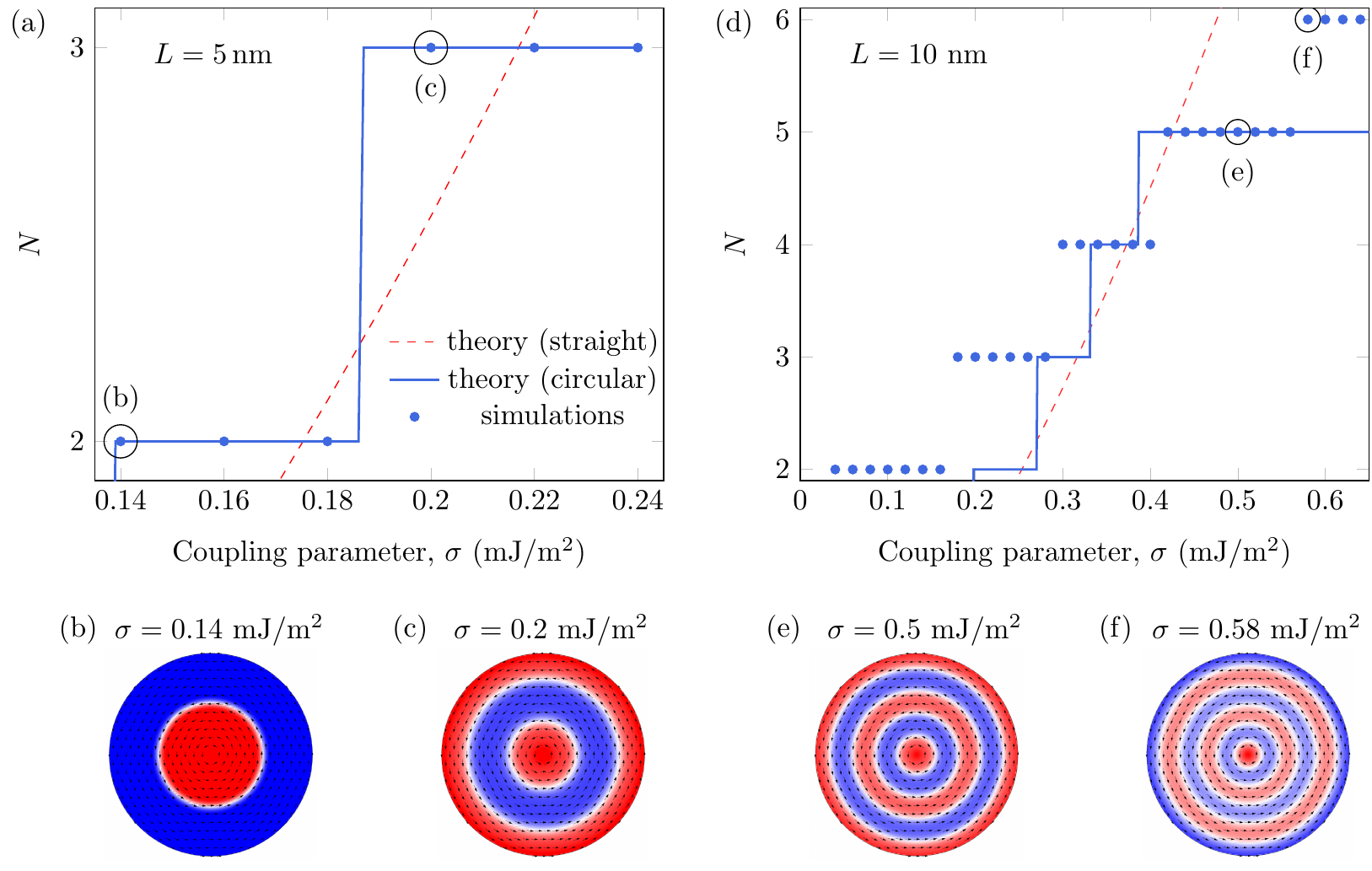}
\end{center}
\caption{\textbf{Number of circular stripe domains} depending on the thickness of the out-of-plane magnetized layer $L$ and the coupling parameter $\sigma$. Results of simulations (symbols) of stripe domains in disks of different thickness: (a) 5-nm-thick and (d) 10-nm-thick disks. Other parameters are the same as in \ref{appendix:oommf_parameters}. Red dashed lines correspond to theoretical dependencies \eqref{eq:nu-linear} for straight stripe domains in infinite samples. Solid blue lines are derived for finite circular stripe domains, see \eqref{eq:N-analytical}. The equilibrium magnetization textures are shown for specified parameters: (b) two-domain state for $L=5~\unit{nm}$ and $\sigma=0.14~\unit{mJ/m^2}$, (c) three-domain state for $L=5~\unit{nm}$ and $\sigma=0.2~\unit{mJ/m^2}$, (e) five-domain state for $L=10~\unit{nm}$ and $\sigma=0.5~\unit{mJ/m^2}$, (f) six-domain state for $L=10~\unit{nm}$ and $\sigma=0.58~\unit{mJ/m^2}$. 
\label{fig:N}%
}
\end{figure*}
%%==================================================================/

We estimate the number of circular stripe domains using the criterion of a modulation instability of a higher symmetry state with respect to the excitation of linear modes. Limiting our consideration by radially symmetric modes with the wave number $k$, we use the boundary conditions $k R=j_{0,\eta}$. Here, $j_{0,\eta}$ defines the continuous generalization of the zeros of the Bessel function enumerated by a positive real number $\eta$, see \ref{appendix:circular} for details.  The critical wave number is provided by the relation
\begin{equation} \label{eq:x-eta}
\frac{R}{L} F(\zeta) = j_{0,\eta}
\end{equation}
with the function $F(\zeta$) given in~\eqref{eq:critical-wave-vector} and parameter $\zeta = {h^2L^2}/{[2\ell^2(Q-1)]}$ for $h>1$ and $\zeta = {L^2}/{[\ell^2(Q-1)]}$ otherwise. Then the number of domains $N$ reads
\begin{equation} \label{eq:N-analytical}
N(\zeta) = \left\lfloor \eta(\zeta) \right\rceil,% \qquad \zeta = \frac{h^2L^2}{2\ell^2(Q-1)},
\end{equation}
where $\lfloor \eta(\zeta) \rceil$ defines the integer closest to $\zeta$ \cite{NIST10}. A comparison with the simulation data is presented in Fig.~\ref{fig:N} revealing a quantitative agreement for small $L$ (Fig.~\ref{fig:N}(a)) and a qualitative agreement for thicker samples, where the edge effect is more significant (Fig.~\ref{fig:N}(b)). The number of stripes $N$ depends on the geometrical and material parameters and increases with the coupling parameter $\sigma$, see Fig.~\ref{fig:N}.

Circular stripe domains are topologically protected textures. Similar to the vortex state \cite{Hubert09}, circular stripes are characterized by the unit $\pi_1$ topological charge, the vorticity. Besides, these states are also characterized by the circulation (clockwise or counter-clockwise). The magnetic texture of circular stripe domains is in many respects similar to another topological texture, skyrmionium \cite{Finazzi13, Komineas15c} or target skyrmions \cite{Leonov14a, Zheng17a, Kent19}. Both textures consist of concentric rings with different values of the out-of-plane magnetization $m_z$, while the in-plane magnetization have specified circulation. The main difference is that the $m_z$ component of skyrmionium changes from $m_z=-1$ to $m_z=+1$, resulting in specific topological properties. Namely, when the number of rings is odd, the $\pi_2$ topological charge (`skyrmion number') is unit, $\mathcal{N}_{\text{sk}}=1$, while for even number of rings $\mathcal{N}_{\text{sk}}=0$. Instead, $\pi_2$ topological properties of the circular stripe domains are the same as for the cone state vortex, namely $\mathcal{N}_{\text{sk}}=q(p-\sqrt{1-\mu^2})/2$. 

%%%%%%%%%%%%%%%%%%%%%%%%%%%%%%%%%%%%%%%%%%%%%%%%%%
%%%			New subsection
%%%%%%%%%%%%%%%%%%%%%%%%%%%%%%%%%%%%%%%%%%%%%%%%%%

\section{Conclusion}
\label{sec:conclusion}

In conclusion, we studied magnetic states in coupled heterostructures consisting of in-plane and out-of-plane magnetized layers separated by a nonmagnetic spacer. If the in-plane magnetized layer hosts magnetic vortex, this texture can modify the magnetic state of the out-of-plane layer leading to the stabilization of topologically protected domain patterns. We demonstrate that by tuning the interlayer exchange coupling, the out-of-plane magnetized layer can accommodate four different states: homogeneous state, cone state vortex, imprinted vortex state, and circular stripe domains. These states were observed experimentally in a heterostructure \ch{Py}/\ch{Pd}/[\ch{Co}/\ch{Pd}] \cite{Streubel15, Streubel15d}. With our work, we provide the fundamental understanding of relevant magnetic interactions, which are responsible for the formation of these non-collinear magnetic states. In particular, we demonstrate that nonlocal magnetostatics cannot be neglected in the model as this is the driving force behind the experimentally visualized circular stripe domains. This state is not stable if only local interactions are taken into account. We discuss similarities and differences of the circular stripe domains with the skyrmionium state. The possibility to realize tunable topologically protected states in coupled heterostructures is potentially relevant for prospective spintronic and spinorbitronic devices relying on chiral non-collinear magnetic textures.  

\ack
We thank Prof. Robert Streubel (University Nebraska-Lincoln), Prof. Oliver G. Schmidt (Technical University of Chemnitz) and Prof. Peter Fischer (Lawrence Berkeley National Laboratory) for their involvement at the initial stages of this project. O.Z. acknowledges support from the UKRATOP project (funded by BMBF under reference 01DK18002). The simulation results were made at the computing cluster of Taras Shevchenko National University of Kyiv \cite{unicc}. This work is financed in part via the German Research Foundation (DFG) under Grant No. MC 9/22–1, MA 5144/24-1, MA 5144/22-1, MA 5144/14-1. 
\appendix

%%%%%%%%%%%%%%%%%%%%%%%%%%%%%%%%%%%%%%%%%%%%%%%%%%
%%%			New section
%%%%%%%%%%%%%%%%%%%%%%%%%%%%%%%%%%%%%%%%%%%%%%%%%%

\section{Variational model of a cone state vortex}
\label{sec:cone-state-appendix}

Light and heavy vortices are known in easy-plane magnets exposed to out-of-plane magnetic fields \cite{Ivanov95b}. They can be excited also in nanodisks \cite{Okuno02, Kravchuk07a}. In a heterostructure with an out-of-plane magnetized layer, vortices can be induced by specific magnetic field of vortex structure. The main difference between the cone state vortices in magnets with in-plane anisotropy and vortices in materials with an out-of-plane easy axis can be demonstrated for magnetic rings: (i) A pure planar vortex forms a ground state of a ring with easy-plane anisotropy without magnetic field. When exposed to an out-of-plane magnetic field, the ground state becomes tilted in the direction of the field forming a light vortex in the cone state. The oppositely directed heavy vortex is unstable in a ring geometry. (ii) Two uniform states with $m_z=\pm1$ form the ground state of a ring-shaped material with the easy-axis anisotropy. When exposed to an in-plane vortex-type magnetic field, the uniform state is not favourable and two cone states with $m_z=\pm\sqrt{1-h^2}$ are realized. As there is no vortex core, both states are energetically equivalent. For the case of a disk of easy-plane magnet, both types of vortices can be realized. But the vortex with the core magnetization oriented along the field direction becomes energetically preferable (mainly due to the magnetization far from the vortex core). The energetically preferred vortex is referred to as the light vortex. In the case of a disk of easy-axis magnet, both types of vortices can coexist in the heterostructure. Nevertheless, one of the vortices has lower energy (mainly due to the vortex core). Similar to the case of an easy-plane magnet, we refer to this vortex as a light vortex. The vortex state with higher energy is referred to as the heavy vortex. 

%%==================================================================/
\begin{figure}
\includegraphics[width=\columnwidth]{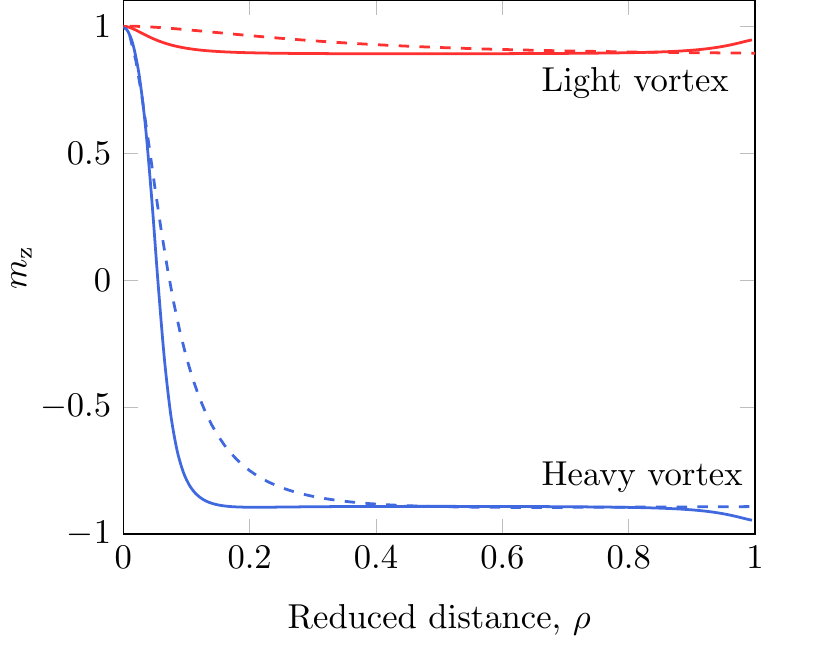}
\caption{\textbf{Light and heavy vortices}: Solid lines correspond to micromagnetic simulations for the sample with a 5-nm-thick out-of-plane magnetized layer for $h=0.47$. Other parameters are the same as in \ref{appendix:oommf_parameters}. Dashed lines describe numerical solutions of the equation \eqref{eq:static}.
\label{fig:light_vs_heavy}%
}
\end{figure}
%%==================================================================/

In the following, we present an analytical model for the description of the cone state vortex. We start from a local model with the energy density \eqref{eq:energy}
\begin{equation} \label{eq:energy-loc}
\mathscr{E}_{\text{loc}} = \lambda^2 \left({\theta'}^2 + \frac{\sin^2\theta}{\rho^2}\right)	+\left(\sin\theta - h\right)^2.
\end{equation}
The structure of the solutions can be found by numerical integration of the boundary value problem:
\begin{equation} \label{eq:static}
\eqalign{&\theta'' + \frac{\theta'}{\rho}-\sin\theta\cos\theta \left(\frac{1}{\lambda^2} + \frac{1}{\rho^2}\right) + \frac{h}{\lambda} \cos\theta=0,\\
&\theta(0)=0, \qquad \theta(1)=\arcsin{h},}
\end{equation}
see Fig.~\eqref{fig:light_vs_heavy}. Micromagnetic simulations confirm the existence of light and heavy vortices as equilibrium states. The analysis shows that the light vortex is the global minimizer. 

%%==================================================================/
\begin{figure}
\includegraphics[width=\columnwidth]{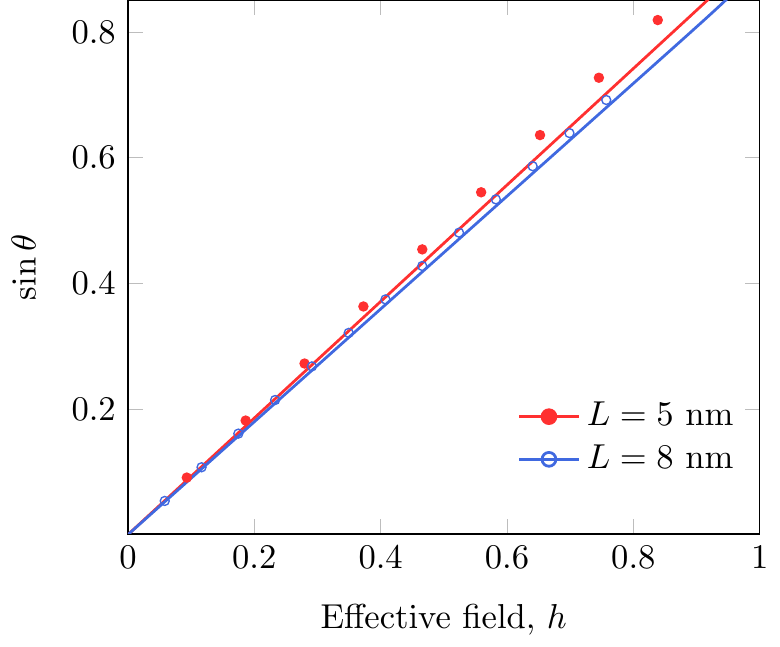}
\caption{\textbf{Vortex shape asymptotic as function of the effective field} for two thicknesses of the layer with out-of-plane easy axis, $L$. Symbols correspond to the results of micromagnetic simulations with parameters as in \ref{appendix:oommf_parameters}. Lines are calculated accordingly to \eqref{eq:ansatz} and \eqref{eq:mu-and-Delta-nloc}.
\label{fig:sintheta_vs_h}%
}
\end{figure}
%%==================================================================/

The analytical description of the light vortex is based on the variational approach using the magnetization profile described by the two-parameter ansatz \eqref{eq:ansatz}. The total energy of the local model reads 
\begin{equation} \label{eq:Eloc}
\eqalign{\mathcal{E}_{\text{loc}} &=\int_0^1 \mathscr{E}_{\text{loc}} \rho  \rmd  \rho \approx \frac{\left(\mu-h\right)^2}{2} + 2\mu h \Delta -\frac{7}{4}\mu^2\Delta^2 \\
&+ \mu^2\lambda^2\left[c_0-\ln\Delta-c_1\ln\left(1-\mu^2\right)\right].}
\end{equation}
Here, $c_0=1/4+\gamma-\ln2\approx 0.134$ with $\gamma\approx 0.577$ being Euler's constant \cite{NIST10}, $c_1\approx 0.09$. The equilibrium values of $\Delta$ and $\mu$ can be found by the minimization of the energy \eqref{eq:Eloc} with respect to these parameters, which results in 
\begin{equation} \label{eq:mu-and-Delta-local}
\eqalign{
\mu_{\text{loc}}&= h\frac{1 - \lambda^2\alpha(h)}{1 -2\lambda^2 \left[\beta(h) +\ln \lambda\right]},\\
\Delta_{\text{loc}}&= \frac{\lambda \sqrt{2}}{\sqrt{\frac{8h}{\mu_{\text{loc}}}-7}}.}	
\end{equation}
The functions $\alpha(h)$ and $\beta(h)$ read:
\begin{equation} \label{eq:alpha-beta}
\eqalign{
\alpha(h) &= 36-\frac{4c_1 h^2 \left(2-h^2\right)}{\left(1-h^2\right)^2},\\
\beta(h) &= c_2 +c_1 \left[\ln\left(1-h^2\right)-\frac{h^2 \left(5-3h^2\right)}{\left(1-h^2\right)^2}\right]
}
\end{equation}
with $c_2=\left(39+\ln2\right)/2-c_0\approx 19.71$. Using typical parameters $\lambda\ll1$, we can limit our consideration to the first terms in a series expansion with respect to $\lambda$, which results in
\begin{equation} \label{eq:mu-and-Delta-local-approx}
\Delta_{\text{loc}}\approx \lambda\sqrt2, \qquad \mu_{\text{loc}}\approx h.	
\end{equation}

%%==================================================================/
\begin{figure}
	\includegraphics[width=\columnwidth]{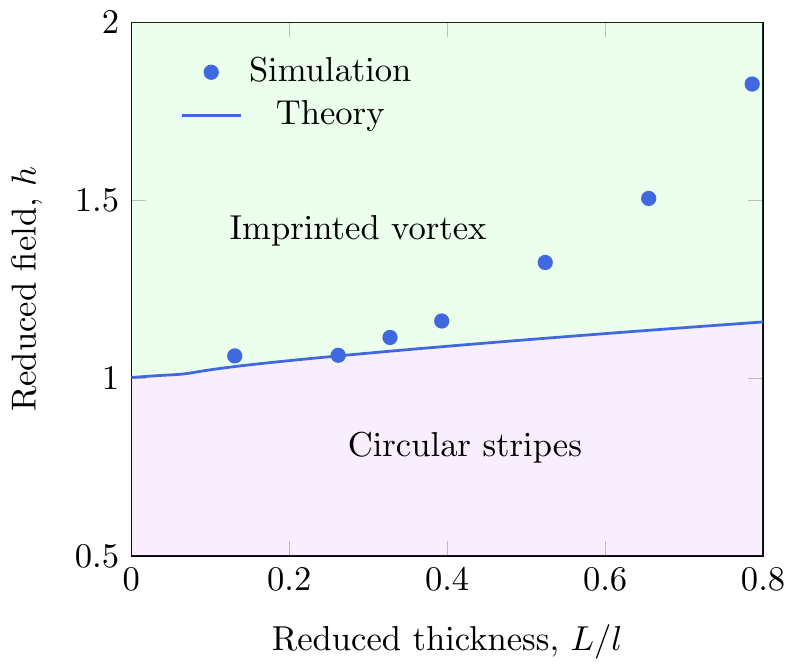}
	\caption{\textbf{Transition to the imprinted vortex:} phase diagram. The boundary between the imprinted vortex and circular stripes is shown with solid line (corresponds to the analytical result \eqref{eq:h-imprinted}) and symbols (corresponds to simulations). Simulation parameters are the same as in \ref{appendix:oommf_parameters}.
		\label{fig:mu1cond_vs_simul}%
	}
\end{figure}
%%==================================================================/

Now we take into account the magnetostatic energy of face surface charges,
\begin{equation} \label{eq:Ems}
E^{\textsc{ms}} = \frac{M_{\textsc{s}}^2}{2} \iint  \frac{m_z(\vec{r})m_z(\vec{r}') \rmd S \rmd S'}{\left|\vec{r} - \vec{r}'\right|}.
\end{equation}
Then the normalized magnetostatic energy reads
\begin{equation} \label{eq:Ems-reduces}
\mathcal{E}^{\textsc{ms}}(\varepsilon) = \frac{1}{2 \varepsilon(Q-1)} \int_0^\infty \left(1-e^{-2\varepsilon \xi}\right) I^2(\xi) \rmd \xi.
\end{equation}
Here, $I(\xi) = \int\limits_0^1 m_z(\rho) \BesselJ_0(\xi\rho) \rho \rmd \rho$, the parameter $\varepsilon = L/(2R)$ is the sample aspect ratio, $Q=K/\left(2\pi M_{\textsc{s}}^2 \right)$ is the quality factor, and $\BesselJ_n(\xi)$ is the Bessel function of the order $n$ \cite{NIST10}. By neglecting a small extra contribution of the exponentially localized vortex core, we limit our consideration by the surface charges of the pure cone state with $m_z=\sqrt{1-\mu^2}$, which results in $I(\xi)=\sqrt{1-\mu^2} \BesselJ_1(\xi)/\xi$. Direct calculations lead to the following expression for the magnetostatic energy:
\begin{subequations} \label{eq:Ems-cone-g}
\begin{eqnarray} \label{eq:Ems-cone}
\!\!&\mathcal{E}^{\textsc{ms}}(\varepsilon) = \frac{\left(1-\mu^2\right)\left[1-g(\varepsilon)\right]}{2(Q-1)},\\
\label{eq:g-vs-epsilon} %
\!\!&g(\varepsilon) = \frac{4}{3\pi}\!\! \left[\! \left(1+\varepsilon^2\right)\ellipticK \! \left(\!\frac{i}{\varepsilon}\!\right)\!\! + \left(1-\varepsilon^2\right)\ellipticE\left(\frac{i}{\varepsilon}\!\right)\!\! -\frac{1}{\varepsilon}\!\right]\!\!,\!\!
\end{eqnarray}	
\end{subequations}
where $\ellipticK(\xi)$ and $\ellipticE(\xi)$ are complete elliptic integrals of the first and second kinds, respectively \cite{NIST10}. Here, $g(\xi)$ is a well localized monotonous function with the limit values $g(0)=0$ and $g(\infty)=1$. The nonlocal contribution to the magnetostatic energy
\begin{equation} \label{eq:Ems-nloc}
\mathcal{E}_{\text{nloc}}^{\textsc{ms}}(\varepsilon) = \mathcal{E}^{\textsc{ms}}(\varepsilon)-\lim_{\varepsilon \to 0}\mathcal{E}^{\textsc{ms}}(\varepsilon) = \frac{\mu^2-1}{2(Q-1)}g(\varepsilon).
\end{equation}
Finally, the total energy of the light vortex $\mathcal{E} = \mathcal{E}_{\text{loc}}$ reads
\begin{equation} \label{eq:Evortex-final}
\eqalign{
\mathcal{E} &= \frac{\left(\mu-h\right)^2}{2} + 2\mu h \Delta +\mu^2\Biggl\{ \frac{g(\varepsilon)}{2(Q-1)} -\frac{7}{4}\Delta\\
& + \lambda^2\left[c_0-\ln\Delta-c_1\ln\left(1-\mu^2\right)\right]\Biggr\},
}	
\end{equation}
where we omitted a constant term. 
Minimization of this energy with respect to $\Delta$ and $\mu$ results in the equilibrium values:
\begin{subequations} \label{eq:mu-and-Delta-nloc}
\begin{eqnarray} \label{eq:Delta-nloc}
\Delta&= \frac{\lambda \sqrt{2}}{\sqrt{\frac{8h}{\mu}-7}}, \\
\label{eq:mu-nloc}
\mu &= h\frac{1 - \lambda^2\alpha(h)}{1 + \frac{g(\varepsilon)}{Q-1} -2\lambda^2 \left[\beta(h) +\ln \lambda\right]}.
\end{eqnarray}	
\end{subequations}
The explicit form of $\alpha(h)$ and $\beta(h)$ is defined in \eqref{eq:alpha-beta}. The function $g(\varepsilon)$ is defined in \eqref{eq:g-vs-epsilon}. In the relevant case of $\lambda\ll1$, we can leave only the first terms in a series expansion with respect to $\lambda$, which results in \eqref{eq:mu-and-Delta}. The amplitude $\mu$ determines the absolute value of the equilibrium in-plane magnetization. According to this expression, far from the vortex core $\sin\theta$ linearly depends on the effective field $h$. Besides, this value varies with the aspect ratio $L/(2R)$ due to the contribution from magnetostatics. This analytical prediction is confirmed by micromagnetic simulations, see Fig.~\ref{fig:sintheta_vs_h}.

The cone state vortex can exist in a specific range of parameters. If the effective field is grater than $h_i$, the imprinted vortex with $m_z=0$ is realized. According to \eqref{eq:mu-nloc}, the critical field reads
\begin{equation} \label{eq:h-imprinted}
\eqalign{
h_i &= \mathfrak{h}(\tilde{\varepsilon}), \qquad \tilde{\varepsilon}=g(\varepsilon),\\
\mathfrak{h}^{-1}(h) &= (Q-1) \Bigl\{h-1 + \lambda^2 \bigl[2\beta(h)\\
&- h\alpha(h) + 2\ln \lambda\bigr]\Bigr\}.\\
}
\end{equation}
We will see below that the same critical picture is valid, when the circular stripe domains are taken into account: the critical field $h_i$ is determined by the condition $\mu=1$. This statement is in a good agreement with simulations for the case of small enough thicknesses, see Fig.~\ref{fig:mu1cond_vs_simul}.

%%%%%%%%%%%%%%%%%%%%%%%%%%%%%%%%%%%%%%%%%%%%%%%%%%
%%%			New section
%%%%%%%%%%%%%%%%%%%%%%%%%%%%%%%%%%%%%%%%%%%%%%%%%%

\section{Straight stripe domains}
\label{sec:straight-stripes}

\begin{figure}
\includegraphics [width=\columnwidth]{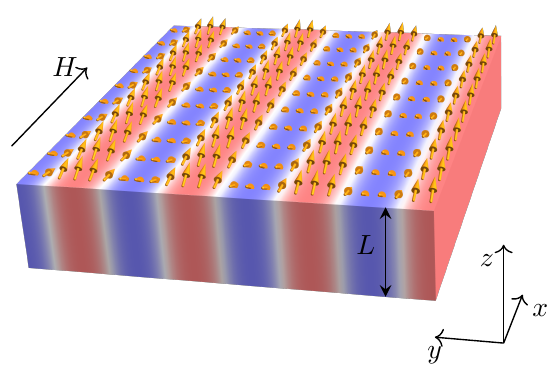}
\caption{ \label{fig:straight}
\textbf{Straight stripe domains.} An extended layer of thickness $L$ with out-of-plane easy axis of magnetization possesses thickness hosts stripe domains whose direction is determined by external uniform maghetic field $H$. Yellow arrows show the local direction of magnetization. Colorcode indicates variation of the out-of-plane component of magnetization.
}
\end{figure}

Here, we consider infinitely thin out-of-plane magnetized film of thickness $L$ exposed to an external in-plane magnetic field $\vec{H}=H\vec{\hat{x}}$, see Fig.~\ref{fig:straight}. We assume that the magnetization depends only on two in-film coordinates $(x,y)$ and suppose that the uniaxial anisotropy $K>2\pi M_{\textsc{s}}^2$. 

%==================================================================/
%==================================================================/

We start with a local model of a thin film described by the energy $E = K_{\text{ef}}L \mathcal{E}$ with $\mathcal{E}= \int \mathscr{E} \rmd S$. The energy density $\mathscr{E}$ reads
\begin{equation} \label{eq:E-film-loc}
\eqalign{\mathscr{E}_{\text{loc}}  &= \ell^2 \left[(\vec{\nabla}\theta)^2 + \sin^2\theta (\vec{\nabla}\phi)^2\right] - \cos^2\theta\\
& - 2\tilde{h}\sin\theta\cos\phi,}
\end{equation}
cf.~\eqref{eq:energy}. Here, $K_{\text{ef}}$ is the effective anisotropy, which takes into account the local part of the magnetostatic interaction (see Sec.~\ref{sec:cone-state}). The reduced magnetic field $\tilde{h}=H/H_0$ with $H_0=2K_{\text{ef}}/M_{\textsc{s}}$. The equilibrium homogeneous state 
$\vec{m}_{0}= \left(\sin\theta_0\cos\phi_0,\sin\theta_0\sin\phi_0, \cos\theta_0 \right)$ is determined similar to \eqref{eq:theta0}:
\begin{equation} \label{eq:m0}
\sin\theta_0=\cases{\tilde{h},& $\tilde{h}< 1$\\1,&$\tilde{h}\ge1$\\}, \qquad \phi=0.
\end{equation}

To describe stripe domains, the nonlocal magnetostatic interaction is essential \cite{Hubert09}. It is convenient to rewrite the normalized magnetostatic energy as $\mathcal{E}^{\textsc{ms}} = -(1/2)\int \vec{m}\cdot \vec{h}^{\textsc{ms}} \rmd V$. The stray field $\vec{h}^{\textsc{ms}} = -\vec{\nabla} \Phi^{\textsc{ms}}$ is caused by the magnetostatic potential of surface and volume charges:
\begin{equation} \label{eq:Phi-ms}
\eqalign{
\Phi^{\textsc{ms}}(\vec{r}) = \frac{1}{4\pi L(Q-1)}&\left[\int \frac{\vec{m}(\vec{r}')\cdot \vec{\hat{n}'}}{|\vec{r}-\vec{r}'|} \rmd S'  \right.\\
&\left.+ \int \frac{\vec{\nabla}'\cdot \vec{m}(\vec{r}')}{|\vec{r}-\vec{r}'|} \rmd V' \right].
}
\end{equation} 

Here, we propose a new approach to analyze the formation of stripe domains in ferromagnetic films with easy-normal anisotropy. 
The periodic domain structure arises as a result of the modulation instability of the equilibrium state $\vec{m}_0$: similar approach was used previously for the description of vortex crystals in Refs.~\cite{Gaididei12a, Kravchuk13}. For this purpose, it is useful to convert the vector Landau-Lifshitz equation $\dot{\vec{m}} = \vec{m}\times \delta{\mathcal{E}}/\delta\vec{m}$ into a scalar form of the Schr\"odinger equation \cite{Gaididei12a, Kravchuk13}:
\begin{equation} \label{eq:Schroedinger}
-i \dot{\psi}=\frac{\delta\mathcal{E}}{\delta\psi^*},
\end{equation} 
where the overdot indicates the derivative with respect to rescaled time $\tau = \omega_0 t$ with $\omega_0 = \gamma_0 K_{\text{ef}}/M_{\textsc{s}}$ and $\gamma_0$ is the gyromagnetic ratio.

Such a transformation can be done using the Holstein--Primakoff--Tyablikov representation \cite{Holstein40,Tyablikov75} of the arbitrary magnetization vector
\begin{equation}\label{eq:HPT}
\eqalign{
\vec{m}&=\vec{m}_0\frac{1-|\psi|^2}{2}+\vec{A}\psi\sqrt{2-|\psi|^2} + \text{c.c.},\\
\vec{A}&=\left(A_x, A_y,A_z\right),\\
A_x &= \frac12 \left(\cos\theta_0\cos\phi_0+i\sin\phi_0 \right),\\
A_y &= \frac12 \left(\cos\theta_0\sin\phi_0 -i\cos\phi_0\right),\\
A_z &= -\frac12\sin\theta_0.
}
\end{equation}
Since Eq.~\eqref{eq:Schroedinger} describes the deviation from a stationary
solution, it has a very convenient form for the analysis of stability
of the given stationary state. For this purpose, we limit our consideration to the harmonic approximation in the energy, taking into account also the magnetostatic energy. In addition, we proceed to the wave-vector space by using the two-dimensional Fourier transform
\begin{equation}
\hat{\psi}_{\vec{k}}=\frac{1}{2\pi} \int_{\mathbb{R}^2}\! \psi(\vec{\rho})e^{-i\vec{\rho}\vec{k}}\rmd \vec{\rho}, \;
\psi(\vec{\rho})=\frac{1}{2\pi} \int_{\mathbb{R}^2}\! \hat{\psi}_{\vec{k}} e^{i\vec{\rho}\vec{k}}\rmd \vec{k}
\end{equation}
with the orthogonality condition
\begin{equation}
\int_{\mathbb{R}^2} e^{i\vec{\rho}(\vec{k}-\vec{k}')}\rmd \vec{\rho}=(2\pi)^2\delta(\vec{k}-\vec{k}')
\end{equation}
where $\vec{\rho}=x\hat{\vec{x}}+y\hat{\vec{y}}$ and $\vec{k}=k_x\hat{\vec{x}}+k_y\hat{\vec{y}}$. The normalized energy $\mathcal{E}$ can be presented as $\mathcal{E}=\int\limits_{\mathbb{R}^2} \hat{\mathscr{E}} \rmd  \vec{k}$, with $\hat{\mathscr{E}}$ being the spectral energy density. Cumbersome but direct calculations result in a spectral energy density in the harmonic approximation:
\begin{equation} \label{eq:E-harmonic}
\eqalign{
\hat{\mathscr{E}} &= \hat{\mathscr{E}}_0 + \mathcal{A}(\vec{k}) |\hat{\psi}_{\vec{k}}|^2 + \frac{1}{2} \left[\mathcal{B}(\vec{k}) \hat{\psi}_{\vec{k}} \hat{\psi}_{-\vec{k}} + \text{c.c.}\right],\\
\mathcal{A}(\vec{k}) &=\ell^2 k^2 + \tilde{h}\sin\theta_0 - \frac{1}{2} \left(1-3\cos^2\theta_0\right) \\
& + G(kL)\left[1-\sin^2\theta_0 \left(1+\frac{k_x^2}{k^2}\right)\right],\\
\mathcal{B}(\vec{k})&= G(kL) \left[\frac{(k_x\cos\theta_0 + ik_y)^2}{k^2}- \sin^2\theta_0\right]\\
& - \frac{\sin^2\theta_0}{2},
}	
\end{equation}
where the function $G(\xi)$ is defined as 
\begin{equation} \label{eq:G(x)}
	G(\xi) = \frac{\xi-1+e^{-\xi}}{2(Q-1)\xi}.
\end{equation}
By neglecting all nonlinear terms in Eq.~\eqref{eq:Schroedinger}, we rewrite equations for complex amplitudes $\hat{\psi}_{\vec{k}}$ and $\hat{\psi}^*_{-\vec{k}}$ as a set of two linear equations
\begin{equation} \label{eq:linear}
-i\dot{\hat{\psi}}_{\vec{k}}=\frac{\delta \mathcal{E}}{\delta\hat{\psi}^*_{\vec{k}}},\qquad i\dot{\hat{\psi}}^*_{-\vec{k}} =\frac{\delta \mathcal{E}}{\delta\hat{\psi}_{-\vec{k}}}.
\end{equation}
Equations \eqref{eq:linear} have the solution $\hat{\psi}_{\vec{k}}= \varPsi_+e^{\mathcal{Z}_+(\vec{k})\tau}$ and $\hat{\psi}^*_{-\vec{k}}=\varPsi_- e^{\mathcal{Z}_-(\vec{k})\tau}$, where 
$\varPsi_\pm(\vec{k})$ are time-independent amplitudes and $\mathcal{Z}_\pm= \pm \sqrt{|\mathcal{B}|^2 - \mathcal{A}^2}$. The last expression determines the stability conditions for the homogeneous equilibrium state $\vec{m}_0$. For $|\mathcal{B}|< \mathcal{A}$, there is a homogeneous equilibrium state. In the opposite case $|\mathcal{B}|> \mathcal{A}$, the equilibrium state $\vec{m}_0$ becomes linearly unstable with respect to modes $\hat{\psi}_{\vec{k}}$. In particular, for the case $k_x=0$, which corresponds to stripe domains oriented along the field, the uniform state becomes unstable when $\mathfrak{F}(q)<0$, where $q=kL$ is the dimensionless wave-vector and
\begin{equation} \label{eq:F(q)}
\mathfrak{F}(q)=\cases{1-\tilde{h}^2+\frac{q^2\ell^2}{L^2}-2\tilde{h}^2 G(q), &  $\tilde{h}<1$,\\\tilde{h}-1+\frac{q^2\ell^2}{L^2}-2 G(q), & $\tilde{h}>1$.\\}
\end{equation}

%%==================================================================/
\begin{figure}
	\includegraphics[width=\columnwidth]{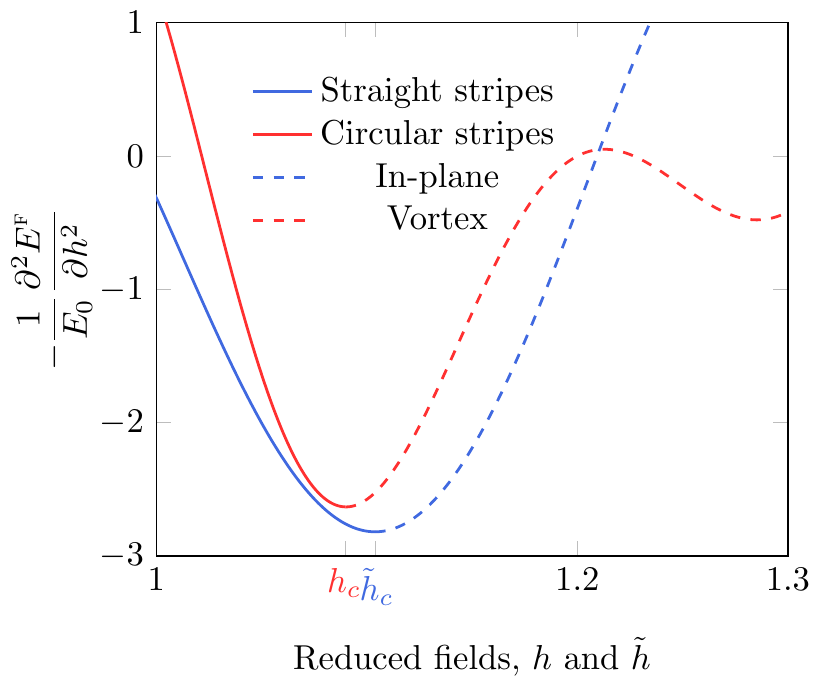}
	\caption{\textbf{Reduced  magnetic susceptibility}: blue lines correspond to the results of micromagnetic simulations of a square-shaped sample with a lateral size of $1 \times 1 $\,$\mu$m$^2$ and periodic boundary conditions. The thickness of the out-of-plane magnetized layer $L=5$~\unit{nm}. Red lines correspond to the results of micromagnetic simulations of a disk-shaped heterostructure with $L=5$\,\unit{nm}. Differential susceptibilities for $h<h_c$ and $\tilde{h}<\tilde{h}_c$ ($h>h_c$ and $\tilde{h}>\tilde{h}_c$) are shown by solid (dashed) lines, respectively. Other parameters are the same as in \ref{appendix:oommf_parameters}. 
	\label{fig:chi}%
	}
\end{figure}
%%==================================================================/

It is instructive to discuss how to separate numerically the stripe domain phase and the in-plane phase. The $\hat{\vec{z}}$-component of magnetization vanishes in both phases near the boundary, that is why we need to identify a quantity which behaves qualitatively different in these two phases. We use the reduced magnetic differential susceptibility $\chi = -(1/E_{0}){\partial^2 E^{\textsc{f}}}/{\partial \tilde{h}^2}$ as such quantity with $E_{0}$ being the energy in the absence of the field and $E^{\textsc{f}}$ being the energy of the interaction with the magnetic field in the case of straight stripes and effective field in the case of circular stripe domains in coupled heterostructures. In the saturated in-plane state $\chi=0$. The stripe state is characterized by a finite value of $\chi$. We define the critical field value $\tilde{h}_c$ as a minimum of the reduced susceptibility, see blue line in Fig.~\ref{fig:chi}. Similar approach we used to separate circular stripe domains and the vortex state in a coupled heterostructure, see red line in Fig.~\ref{fig:chi}.

%==================================================================/
\begin{figure}
\includegraphics[width=\columnwidth]{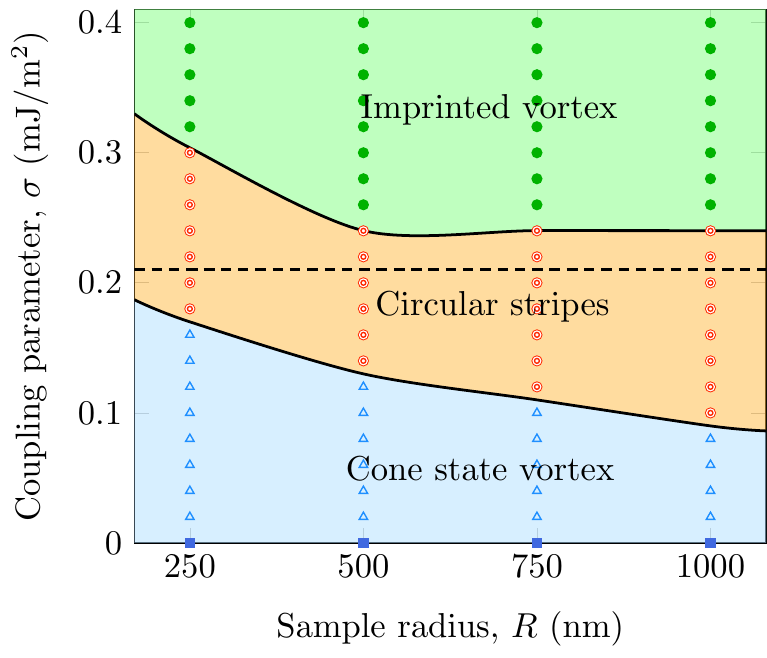}
\caption{\textbf{Phase diagram for heterostructures shaped as disks of different radii}: micromagnetic simulations for a disk-shaped heterostructure containing a 5-nm-thick out-of-plane magnetized layer. Parameters are the same as in \ref{appendix:oommf_parameters}. Dashed line describes the boundary of linear instability of the cone state vortex.
}\label{fig:sigmaR}
\end{figure}
%==================================================================/

We performed micromagnetic simulations for samples of different radii, see Fig.~\ref{fig:sigmaR}. A general trend is as follows: the coupling strength, which is needed to realize circular stripe domains, decreases with the sample radius. At the same time, the boundary of the linear instability of the cone state vortex is almost independent of the sample radius. The number of circular stripe domains increases with the system radius in a qualitative agreement with \eqref{eq:N-analytical}.

%%==================================================================/
\begin{figure}
\includegraphics[width=\columnwidth]{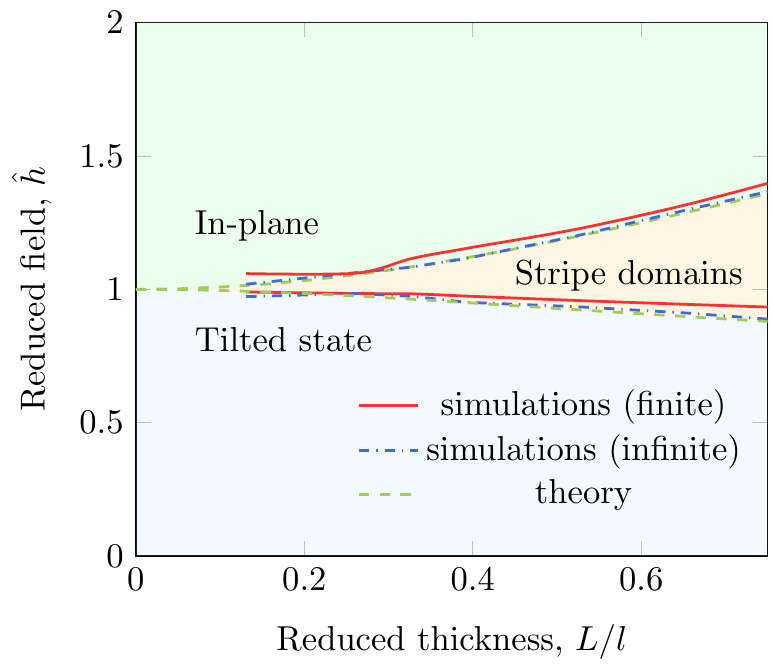}
\caption{\textbf{Straight stripe domains}: phase diagram. Lines describe boundaries between different phases: (i) Solid red lines correspond to the results of micromagnetic simulations of a square-shaped film with lateral dimensions of $1 \times 1 $\,$\mu$m$^2$. (ii) Dash-dotted blue lines correspond to the results of micromagnetic simulations of the same square sample with periodic boundary conditions. This case mimics an infinite system, which is analysed analytically. (iii) Dashed green lines correspond to the theoretically obtained boundary accordingly to \eqref{eq:F(q)}.
\label{fig:straight_stripes}%
}
\end{figure}
%%==================================================================/

When studying straight stripe domains, we determined that the uniform state becomes linearly unstable when $\mathfrak{F}(q)<0$, see Eq.~\eqref{eq:F(q)}. In the following, we analyse consequences of this instability. The dependence $\mathfrak{F}(q)$ is a nonmonotonic one. Hence, we can determine the critical parameters, when the instability starts to develop, from the condition $\mathfrak{F}'(q_c)=0$. The critical wave-vector $q_c=F(\zeta)$, where $\zeta=\frac{\sin^2\theta_0 L^2}{2\ell^2(Q-1)}$, $\theta_0$ is determined by \eqref{eq:m0}, and the inverse function $\zeta=F^{-1}(q_c)$ is determined as
\begin{equation} \label{eq:critical-wave-vector}
F^{-1}(q) = \frac{q^3}{1-e^{-q}(1+q)}.
\end{equation}
The function $F(\zeta)$ has the asymptotic $F(\zeta)\approx \zeta/2$ when $\zeta\ll1$ and $F(\zeta)\approx \sqrt[3]{\zeta}$ when $\zeta\gg1$.

The wave-vector $q=kL$ determines the linear density of domains. Using the periodic boundary conditions
\begin{equation} \label{eq:BC-straight}
k L_y=\pi N	
\end{equation}
with $L_y$ being the linear film size in $\hat{\vec{y}}$-direction, we can define the reduced density of domains $\nu=N L_x/L_y$ as
\begin{equation} \label{eq:nu-linear}
\nu = \frac{q}{\pi}	 = \frac{1}{\pi} F\left(\frac{\sin^2\theta_0 L^2}{2\ell^2(Q-1)}\right).
\end{equation}
The density of domains has the following asymptotic dependence on the film thickness:
\begin{equation} \label{eq:nu-linear-asymptote}
\nu\approx \cases{
\frac{\sin^2\theta_0 L^2}{4\pi \ell^2(Q-1)},& when $L\ll \frac{\ell\sqrt{Q-1}}{\sin\theta_0}$\\
\sqrt[3]{\frac{\sin^2\theta_0 L^2}{2\pi^3\ell^2(Q-1)}},& when $L\gg \frac{\ell\sqrt{Q-1}}{\sin\theta_0}$.\\}	
\end{equation}

To verify these analytical predictions, we performed full scale micromagnetic simulations of a square-shaped sample with a lateral size of $1 \times 1 $\,$\mu$m$^2$, see Fig.~\ref{fig:straight_stripes}. The corresponding theoretical curves are shown by dashed green lines. The observed discrepancies between the theoretical results and simulations data are caused by the fact that the theory was constructed for an infinitely extended film. To model infinite system, we performed simulations with the same square sample under the periodic boundary conditions in $\hat{\vec{x}}$ and $\hat{\vec{y}}$, see dash-dotted blue lines. In this case, we obtain a very accurate correspondence with the theoretical predictions.

\section{Number of domains: straight vs circular stripes}
\label{appendix:circular}

%%==================================================================/
\begin{figure}
\includegraphics[width=\columnwidth]{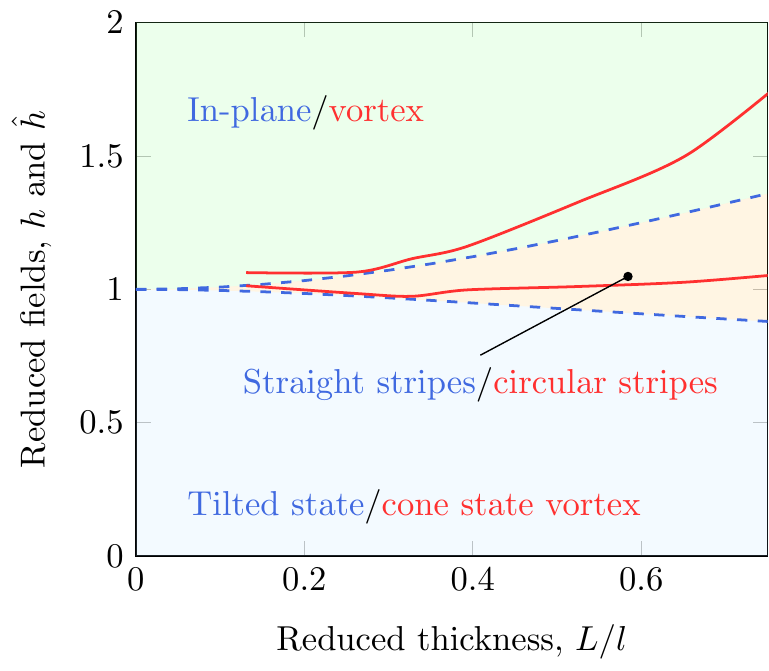}
\caption{\textbf{Straight vs circular stripe domains}: Solid red lines correspond to the results of micromagnetic simulations of boundaries between different phases in a disk-shaped heterostructure containing a 5-nm-thick out-of-plane magnetized layer. Parameters are the same as in \ref{appendix:oommf_parameters}. Dashed blue lines are critical instability curves calculated accordingly to \eqref{eq:F(q)} within the theory of straight stripe domains in an infinite film.
\label{fig:straight_vs_circular}%
}
\end{figure}
%%==================================================================/

%%==================================================================/
\begin{figure}
\includegraphics[width=\columnwidth]{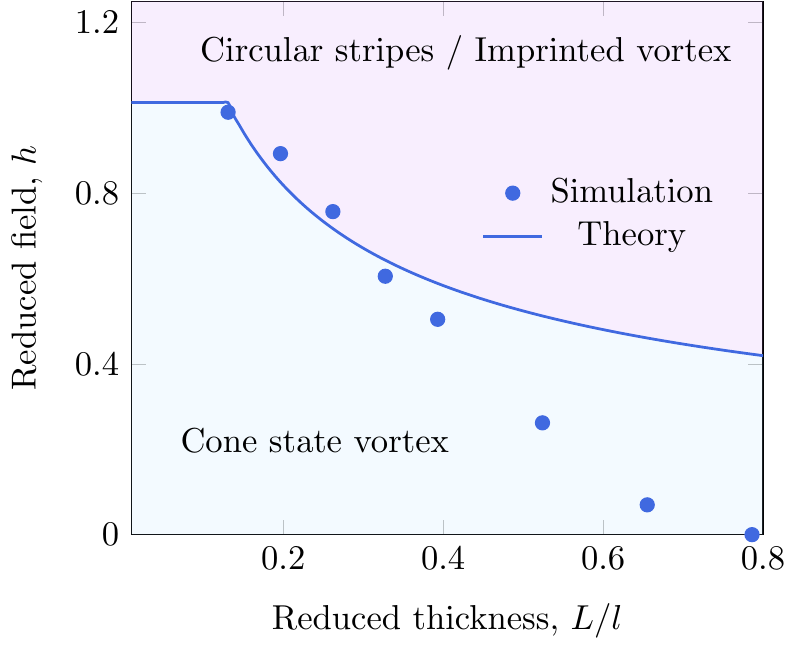}
\caption{\textbf{Transition to circular stripe domains:} phase diagram. The boundary between the cone state vortex and circular stripe domains is shown with solid line (corresponds to numerical calculations of \eqref{eq:x-1_5}) and symbols (corresponds to simulations). Simulation parameters are the same as in \ref{appendix:oommf_parameters}.
\label{fig:circular_bottom_theoryvssimulation}%
}
\end{figure}
%%==================================================================/

In the following, we apply the above analysis for the straight stripe domains in infinite films to describe the circular stripe domain pattern in disk-shaped coupled heterostructures. For this purpose, we compare results of full scale micromagnetic simulations for the heterostructure (solid red lines in Fig.~\ref{fig:straight_vs_circular}) with the theoretical analysis of instabilities for the case of straight stripe domains (dashed blue lines in Fig.~\ref{fig:straight_vs_circular}). The results of simulations and theory agree for relatively thin samples. The discrepancy for thicker samples has several reasons: the influence of the confinement, edge effects, and the interfacial nature of the interlayer exchange coupling. These effects should be studied separately and are not addressed in this work.

To address the number of circular domains, we use an approach, which is similar to procedure described in \ref{sec:straight-stripes}. Presence of magnetostatics complicates the boundary conditions. This can be taken into account by the effective boundary pinning of magnetization~\cite{Guslienko02b}. For a static problem in cylindrical geometry, it is convenient to find the radially symmetric solution as a linear combination of Bessel and Neumann functions which should be zero at the sample's boundary. Then, the wave number $k$ can be found from the condition $kR = j_{0,\eta}$, where $\eta$ is a positive real number and $j_{0,\eta}$ is a solution of
\begin{equation} \label{eq:cylindricals:olek}
 \BesselJ_0(j_{0,\eta})\cos\left(\pi\{\eta\}\right) = \BesselY_0(j_{0,\eta})\sin\left(\pi\{\eta\}\right),
\end{equation}
c.f. Eq.~\eqref{eq:BC-straight}. Here, expressions $\lfloor \eta \rceil$ and $\{\eta\}$ denote the integer closest to $\eta$ and the fractional part of $\eta$, respectively, and $\BesselY_0(\bullet)$ is the Neumann function~\cite{NIST10}. Then, the reduced wave number reads
\begin{equation} \label{eq:q-eta}
q=\frac{L}{R}j_{0,\eta}=F(\zeta)	
\end{equation}
with $F(\zeta)$ being implicitly defined by \eqref{eq:critical-wave-vector}. Finally, this results in the number of circular stripe domains \eqref{eq:N-analytical}.

We apply the above theoretical analysis to describe the boundary between the cone state vortex and circular stripe domains. Using the condition
\begin{equation} \label{eq:x-1_5}
\frac{R}{L} F(\zeta) = j_{0,\eta^\star}, \quad \zeta = \frac{h^2L^2}{2\ell^2(Q-1)},
\end{equation}
with $\eta^\star=1.5$ \cite{Note3}, we identify the critical curve $h(L)$, see Fig.~\ref{fig:circular_bottom_theoryvssimulation}, which corresponds to the full scale micromagnetic simulations for thin samples.

\section{Parameters used for OOMMF simulations}
\label{appendix:oommf_parameters}

In-plane magnetized disk of radius $R = 500\,\unit{nm}$ with magnetic parameters of Permalloy (\ch{Py}, \ch{\py}; exchange constant $A=13~\unit{pJ/m}$; saturation magnetization $M_\textsc{s} = 860~\unit{KA/m}$). Out-of-plane magnetized disk of radius $R = 500 \unit{nm}$ and thickness $L$ with parameters typical for \ch{Co/Pt} multilayers  ($A=10~\unit{pJ/m}$, $M_\textsc{s} = 500~\unit{kA/m}$, $K = 200~\unit{kJ/m^3}$)~\cite{Streubel15d}. These two magnetic layers are separated by a nonmagnetic spacer of thickness $d$. Thermal effects are neglected in simulations. Mesh cells have size of $5\times 5\times 2~\unit{nm}^3$.

%%%%%%%%%%%%%%%%%%%%%%%%%%%%%%%%%%%%%%%%%%%%%%%%%%
%%%		Bibliography
%%%%%%%%%%%%%%%%%%%%%%%%%%%%%%%%%%%%%%%%%%%%%%%%%%
%\bibliography{soliton,xtra}

\begin{thebibliography}{10}
	\expandafter\ifx\csname url\endcsname\relax
	\def\url#1{{\tt #1}}\fi
	\expandafter\ifx\csname urlprefix\endcsname\relax\def\urlprefix{URL }\fi
	\providecommand{\eprint}[2][]{\url{#2}}
	
	\bibitem{Parkin15}
	Parkin S and Yang S~H 2015 {\em Nature Nanotechnology\/} {\bf 10} 195--198
	\urlprefix\url{http://dx.doi.org/10.1038/nnano.2015.41}
	
	\bibitem{Zang18}
	Zang J, Cros V and Hoffmann A (eds) 2018 {\em Topology in Magnetism\/}
	(Springer International Publishing)
	
	\bibitem{Grollier20}
	Grollier J, Querlioz D, Camsari K~Y, Everschor-Sitte K, Fukami S and Stiles M~D
	2020 {\em Nature Electronics\/} {\bf 3} 360--370
	
	\bibitem{Goebel21}
	G\"{o}bel B, Mertig I and Tretiakov O~A 2021 {\em Physics Reports\/} {\bf 895}
	1--28
	
	\bibitem{Fert17}
	Fert A, Reyren N and Cros V 2017 {\em Nature Reviews Materials\/} {\bf 2} 17031
	\urlprefix\url{https://doi.org/10.1038/natrevmats.2017.31}
	
	\bibitem{Wiesendanger16}
	Wiesendanger R 2016 {\em Nature Reviews Materials\/} {\bf 1} 16044
	\urlprefix\url{http://dx.doi.org/10.1038/natrevmats.2016.44}
	
	\bibitem{Jiang15}
	Jiang W, Upadhyaya P, Zhang W, Yu G, Jungfleisch M~B, Fradin F~Y, Pearson J~E,
	Tserkovnyak Y, Wang K~L, Heinonen O, te~Velthuis S~G~E and Hoffmann A 2015
	{\em Science\/} {\bf 349} 283--286
	\urlprefix\url{http://dx.doi.org/10.1126/science.aaa1442}
	
	\bibitem{Streubel15}
	Streubel R, Han L, Im M~Y, Kronast F, R{\"o\ss}ler U~K, Radu F, Abrudan R, Lin
	G, Schmidt O~G, Fischer P and Makarov D 2015 {\em Scientific Reports\/} {\bf
		5} 8787 ISSN 2045-2322 \urlprefix\url{http://dx.doi.org/10.1038/srep08787}
	
	\bibitem{Streubel15d}
	Streubel R, Fischer P, Kopte M, Schmidt O~G and Makarov D 2015 {\em Applied
		Physics Letters\/} {\bf 107} 112406
	\urlprefix\url{http://dx.doi.org/10.1063/1.4931101}
	
	\bibitem{Hubert09}
	Hubert A and Sch{\"a}fer R 2009 {\em {Magnetic domains: The analysis of
			magnetic microstructures}\/} (Berlin: Springer Berlin Heidelberg) ISBN
	978-3-540-64108-7
	\urlprefix\url{https://www.ebook.de/de/product/1342062/alex_hubert_rudolf_schaefer_magnetic_domains.html}
	
	\bibitem{Stiles99}
	Stiles M 1999 {\em Journal of Magnetism and Magnetic Materials\/} {\bf 200}
	322--337
	
	\bibitem{OOMMFa}
	The {O}bject {O}riented {M}icro{M}agnetic {F}ramework developed by M. J.
	Donahue and D. Porter mainly, from NIST. We used the 3D version of the 1.2b3
	release \urlprefix\url{http://math.nist.gov/oommf/}
	
	\bibitem{Donahue99}
	Donahue M~J and Porter D~G 1999 {OOMMF} user's guide, version 1.0 Tech. rep.
	Interagency Report NISTIR 6376
	\urlprefix\url{http://math.nist.gov/~MDonahue/pubs/abstracts.html#Donahue199909}
	
	\bibitem{Suna86}
	Suna A 1986 {\em Journal of Applied Physics\/} {\bf 59} 313--316
	(\textit{Preprint} \eprint{https://doi.org/10.1063/1.336684})
	\urlprefix\url{https://doi.org/10.1063/1.336684}
	
	\bibitem{Lemesh17}
	Lemesh I, B\"uttner F and Beach G~S~D 2017 {\em Physical Review B\/} {\bf
		95}(17) 174423
	\urlprefix\url{https://link.aps.org/doi/10.1103/PhysRevB.95.174423}
	
	\bibitem{Kosevich83}
	Kosevich A~M, Voronov V~P and Manzhos I~V 1983 {\em Sov. Phys. JETP\/} {\bf 57}
	148 \urlprefix\url{http://www.jetp.ac.ru/cgi-bin/e/index/r/84/1/p148?a=list}
	
	\bibitem{Ivanov95b}
	Ivanov B~A and Sheka D~D 1995 {\em Low Temperature Physics\/} {\bf 21} 881--887
	\urlprefix\url{http://link.aip.org/link/?LTP/21/881/1}
	
	\bibitem{Feldtkeller65}
	Feldtkeller E and Thomas H 1965 {\em Zeitschrift f{\"u}r Physik B Condensed
		Matter\/} {\bf 4} 8--14 \urlprefix\url{http://dx.doi.org/10.1007/BF02423256}
	
	\bibitem{Note2}
	{The weak dependence of the vortex core size on the field appears due to the
		the field dependence of parameters $\alpha(h)$ and $\beta(h)$, see
		\eqref{eq:alpha-beta}, which is negligible in the range of parameters
		$\lambda\ll1$.}
	
	\bibitem{Thiaville03}
	Thiaville A, Garcia J~M, Dittrich R, Miltat J and Schrefl T 2003 {\em Physical
		Review B\/} {\bf 67} 094410 (pages~12)
	\urlprefix\url{http://link.aps.org/abstract/PRB/v67/e094410}
	
	\bibitem{Kravchuk07a}
	Kravchuk V and Sheka D 2007 {\em Physics of the Solid State\/} {\bf 49}
	1923--1931 \urlprefix\url{http://dx.doi.org/10.1134/S1063783407100186}
	
	\bibitem{NIST10}
	Olver F~W~J, Lozier D~W, Boisvert R~F and Clark C~W (eds) 2010 {\em NIST
		Handbook of Mathematical Functions\/} (New York, NY: Cambridge University
	Press) ISBN 0521140633
	\urlprefix\url{http://www.cambridge.org/us/academic/subjects/mathematics/abstract-analysis/nist-handbook-mathematical-functions}
	
	\bibitem{Finazzi13}
	Finazzi M, Savoini M, Khorsand A~R, Tsukamoto A, Itoh A, Du{\`{o}} L, Kirilyuk
	A, Rasing T and Ezawa M 2013 {\em Physical Review Letters\/} {\bf 110} 177205
	\urlprefix\url{http://dx.doi.org/10.1103/PhysRevLett.110.177205}
	
	\bibitem{Komineas15c}
	Komineas S and Papanicolaou N 2015 {\em Physical Review B\/} {\bf 92}(6) 064412
	\urlprefix\url{http://dx.doi.org/10.1103/PhysRevB.92.064412}
	
	\bibitem{Leonov14a}
	Leonov A~O, R{\"o}{\ss}ler U~K and Mostovoy M 2014 {\em {EPJ} Web of
		Conferences\/} {\bf 75} 05002
	
	\bibitem{Zheng17a}
	Zheng F, Li H, Wang S, Song D, Jin C, Wei W, Kov\'acs A, Zang J, Tian M, Zhang
	Y, Du H and Dunin-Borkowski R~E 2017 {\em Physical Review Letters\/} {\bf
		119}(19) 197205
	\urlprefix\url{https://link.aps.org/doi/10.1103/PhysRevLett.119.197205}
	
	\bibitem{Kent19}
	Kent N, Streubel R, Lambert C~H, Ceballos A, Je S~G, Dhuey S, Im M~Y, Büttner
	F, Hellman F, Salahuddin S and Fischer P 2019 {\em Applied Physics Letters\/}
	{\bf 115} 112404 (\textit{Preprint}
	\eprint{https://doi.org/10.1063/1.5099991})
	\urlprefix\url{https://aip.scitation.org/doi/10.1063/1.5099991}
	
	\bibitem{unicc}
	High--performance computing cluster of {T}aras {S}hevchenko {N}ational
	{U}niversity of {K}yiv \url{http://cluster.univ.kiev.ua/eng/}
	\urlprefix\url{http://cluster.univ.kiev.ua/eng/}
	
	\bibitem{Okuno02}
	Okuno T, Shigeto K, Ono T, Mibu K and Shinjo T 2002 {\em Journal of Magnetism
		and Magnetic Materials\/} {\bf 240} 1--6
	\urlprefix\url{https://www.sciencedirect.com/science/article/abs/pii/S0304885301007089}
	
	\bibitem{Gaididei12a}
	Gaididei Y, Volkov O~M, Kravchuk V~P and Sheka D~D 2012 {\em Physical Review
		B\/} {\bf 86}(14) 144401
	\urlprefix\url{http://link.aps.org/doi/10.1103/PhysRevB.86.144401}
	
	\bibitem{Kravchuk13}
	Kravchuk V~P, Volkov O~M, Sheka D~D and Gaididei Y 2013 {\em Physical Review
		B\/} {\bf 87}(22) 224402
	\urlprefix\url{http://link.aps.org/doi/10.1103/PhysRevB.87.224402}
	
	\bibitem{Holstein40}
	Holstein T and Primakoff H 1940 {\em Physical Review\/} {\bf 58}(12) 1098--1113
	\urlprefix\url{http://link.aps.org/doi/10.1103/PhysRev.58.1098}
	
	\bibitem{Tyablikov75}
	Tyablikov S~V 1975 {\em Methods in the Quantum Theory of Magnetism\/} 2nd ed
	(Moscow, Nauka) [transl. of 1st Russ. ed., Plenum Press, New York (1967)]
	
	\bibitem{Guslienko02b}
	Guslienko K~Y, Demokritov S~O, Hillebrands B and Slavin A~N 2002 {\em Physical
		Review B\/} {\bf 66} 132402
	\urlprefix\url{http://dx.doi.org/10.1103/PhysRevB.66.132402}
	
	\bibitem{Note3}
	{The value $\eta^\star=1.5$ is the smallest value when $\lfloor \eta \rceil =
		2$ and stripe domains become energetically preferable.}
	
\end{thebibliography}
\providecommand{\newblock}{}

\end{document}